\documentstyle[aps,preprint]{revtex}
\tightenlines

\begin{document}
\draft

\title{Scattering functions for multicomponent mixtures of charged hard spheres,
including the polydisperse limit. Analytic expressions in the Mean Spherical
Approximation.}

\author{Domenico Gazzillo}
\address{INFM Unit\`{a} di Venezia e Dipartimento di
Chimica Fisica, Universit\`{a} di Venezia, S. Marta 2137,
I-30123 Venezia, Italy} 
\author{Achille Giacometti}
\address{INFM Unit\`{a} Venezia e Dipartimento di Scienze Ambientali, 
Universit\`{a} di Venezia, S. Marta 2137, I-30123 Venezia, Italy} 
\author{Flavio Carsughi}
\address{INFM Unit\`{a} di Ancona and Institut f\"{u}r
 Festk\"{o}rperforschung des Forschungzentrum J\"{u}lich,
 D-52425 J\"{u}lich, Germany}

\date{\today}
\maketitle

\begin{abstract}
We present a closed analytical formula for the scattering intensity from
charged hard sphere fluids with any arbitrary number of components. Our
result is an extension to ionic systems of Vrij's analogous expression for
uncharged hard sphere mixtures. Use is made of Baxter's factor correlation
functions within the Mean Spherical Approximation (MSA). The polydisperse
case of an infinite number of species with a continuous distribution of hard
sphere diameters and charges is also considered. As an important by-product
of our investigation, we present some properties of a particular kind of
matrices (sum of the identity matrix with a {\it dyadic} matrix)
appearing in the solution of the MSA integral equations for both uncharged
and charged hard sphere mixtures. This analysis provides a general framework
to deal with a wide class of MSA solutions having dyadic structure and
allows an easy extension of our formula for the scattering intensity to
different potential models. Finally, the relevance of our results for the
interpretation of small angle neutron scattering experimental data is
briefly discussed.
\end{abstract}

\newpage
\narrowtext

\section{Introduction}

Most of the naturally occurring or industrially important fluids are
actually mixtures of more than two components. In particular, liquid
mixtures of macroparticles (e.g., micellar or colloidal suspensions) exhibit
size, shape and possibly charge distributions of their components.
Macroparticles, unlike atoms, are intrinsecally polydisperse.

As a consequence, any rigorous statistical mechanical model of real mixtures
should be able to include a large finite number $p$ of species, which
eventually will become infinite in the presence of polydispersity. In the
latter case the properties of the mixtures will depend upon idealized
continuous distributions of size, charge, etc..

Unfortunately, the presence of a large number of components poses
challenging problems to the two major tools of the liquid state theory,
namely computer simulation and integral equation (IE) methods. In fact, in
the first case a prohibitively large number of particles may be required for
a satisfactory Monte Carlo or molecular dynamics simulation, with
consequently long and expensive computational times. In the second case,
when the IEs must be solved numerically (as almost always occurs), the
algorithms work with arrays formed by about $10^3\times p^2$ elements, so
that larger and larger amounts of memory and again of computational time are
required as $p$ increases; moreover, non-convergence problems in the
numerical solution are encountered more frequently, especially in the
presence of strong correlations and high density. Finally, polydisperse
systems with infinitely-many components cannot be investigated in a fully
adequate way if the IEs have to be solved numerically. In such cases, a
discretization is unavoidable and one is forced to replace the continuous
distribution of species with an appropriate $p$-component mixture, as shown
by D'Aguanno et al. \cite{Aguan90}, \cite{Aguan91}, \cite{Wagner91}%
, \cite{Krause91}, \cite{Aguan92} who solved integral equations with $p\leq
10$ for Yukawa ($\equiv \ $screened Coulomb) plus hard core interaction
potentials. However, apart from the aforesaid work, most of the published IE
studies on fluid mixtures concern binary or, more rarely, ternary systems 
\cite{Hoshino83}, \cite{Gazzillo94}, \cite{Gazzillo95}.

In this paper we are interested in the static structural properties of
multicomponent fluids. Experimentally, these can be efficiently monitored
using X-ray, neutron and light scattering. Fitting experimental structure
factors or scattering intensity of systems with a large number of components
using IE theories is actually feasible only if the relevant equations do
admit analytical solutions. Such a fortunate opportunity occurs only for
peculiar interaction potentials and, to the best of our knowledge, this is
limited to a single class of approximate equations.

The simplest case refers to the model mixture with hard sphere (HS)
interactions. Using Baxter's analytical solution of the approximate
Percus-Yevick (PY) integral equation for HS mixtures \cite{Baxter70}, Vrij 
\cite{Vrij79} derived a closed expression for the scattering intensity $R(k)$
which holds true for {\it any} number $p$ of components. It is remarkable
that Vrij's formula depends on $p$ only through the averages of some
quantities. Hence, only the number of terms involved in these averages
increases as $p$ increases.This feature allows to extend the application of
this equation for $R(k)$ to polydisperse HS mixtures \cite{Beurten81}. The
results are in a satisfactory agreement with Monte Carlo simulation data 
\cite{Frenkel86}.

Independently and along a somewhat different route, Blum and Stell \cite
{Stell79} presented the PY scattering function for polydisperse fluids of
hard or permeable spheres in terms of Fourier transforms $H_{ij}(k)$ of the
total correlation functions (some misprints are corrected in Appendix A of
ref. \cite{Mcrae88}). In addition, the averages involved in Blum and Stell's
expressions were evaluated analytically by Griffith et al. \cite
{Griffith86} in the case of polydisperse mixtures with a continuous Schultz
distribution of diameters.

Now, apart from the HS case, we are unaware of a similar analytic
determination of scattering functions for any other interaction potentials 
\cite{Pedersen94}, \cite{Kline96}. The main purpose of the present paper is
then to extend Vrij's work to multicomponent mixtures of charged particles.

The simplest model for charged systems is the so-called {\it primitive
model, }consisting of an electroneutral mixture of charged hard spheres
embedded in a dielectric continuum and interacting through unscreened
Coulomb potentials. For these pair potentials, the unrestricted general
solution, within the {\it Mean Spherical Approximation (MSA),} is
available \cite{Blum75},\cite{Blum77},\cite{Blum80A}, \cite{Blum80B}: simple
analytical expressions for the Baxter factor correlation functions $%
q_{ij}(r) $ \cite{Baxter70} are given and only a single non-linear equation
in one parameter, $\Gamma $, has to be solved.

It is well known that, for dilute solutions of highly charged particles, the
MSA may lead to unphysical negative values for some of the radial
distribution functions $g_{ij}(r)$ close to the contact distance or in a
neighborhood of the first minimum (in fact, the MSA is asymptotically
correct for $r\rightarrow \infty $, but it is incorrect at short distances,
in the region just outside the core). To overcome this drawback and extend
the validity of the MSA to arbitrary low densities, some authors have
proposed an {\it ad hoc} rescaling method, which replaces large
electrostatic repulsions by appropriate HS repulsions \cite{Hansen82},\cite
{Belloni86},\cite{Ruiz90} . Such a ``rescaled mean spherical approximation''
(RMSA) preserves the analytical form of the MSA solution and therefore
allows an iterative fit of low density experimental data.

Nevertheless, in the regime of weakly charged solutions at sufficiently high
concentrations, the MSA is expected to be a reasonably accurate
approximation (in fact, the Coulomb part of the potentials can be considered
as a perturbation with respect to the HS one, whereas in the above-mentioned
opposite regime the electrostatic effects predominate over the HS
repulsions).

The primitive model can be utilized for a large class of ionic fluid
mixtures (electrolyte solutions, molten salts, solutions of macromolecules
and polymers, micellar and colloidal suspensions, microemulsions, etc.) and
with a careful analysis of the regime of validity for each case the MSA can
be safely employed. For instance, Abramo et al. \cite{Abramo78}
applied the MSA analytic solution for charged HS to the evaluation of
partial structure factors and x-ray diffraction patterns for the whole
family of molten alkali halides (binary systems). Caccamo and Malescio \cite
{Caccamo90} compared MSA and HNC (hypernetted-chain) results for structural
and thermodynamic properties of polyelectrolytes (in particular, micellar
solutions with an added electrolyte). We also note that Senatore and Blum 
\cite{Senatore85} already performed MSA calculations of the average
structure factor $S^M(k)$ for mixtures of charged HS with either size
polydispersity or charge polydispersity. However, none of these authors
presented closed-form analytical expressions for the MSA structure functions
as in Vrij's work \cite{Vrij79}; this is the task we have accomplished here.

The paper is organized as follows. In the next Section we briefly review the
basic formalism of scattering and integral equation theory for
multicomponent fluids. In Section III we discuss some useful properties of a
peculiar class of matrices (related to dyadic matrices) encountered in the
MSA solution for both neutral and charged HS mixtures (for uncharged HS the
MSA coincides with the PY approximation). Section IV then provides the
analytical MSA equations required to derive the new formula for the
scattering intensity (and for the average structure factor). This is
described later on in Section V. The polydisperse limit is briefly discussed
(Section VI) and, finally, in Section VII we summarize our results and
comment on perspectives for future investigation.

\section{Scattering and Integral Equation Theory}

\strut

\subsection{Scattering\ functions}

The {\it coherent scattering intensity }$I(k)$ for a $p$-component fluid
mixture with spherically symmetric interparticle interactions can be written
in terms of the partial structure factors $S_{ij}\left( k\right) $ as \cite
{Waseda80}

\begin{equation}
R(k)\equiv I\left( k\right) /V=\sum_{i,j=1}^p\left( \rho _i\rho _j\right)
^{1/2}F_i\left( k\right) F_j\left( k\right) S_{ij}\left( k\right) ={\bf f}%
^T\left( k\right) {\bf S}\left( k\right) {\bf f}\left( k\right) 
\label{f1}
\end{equation}

\noindent Here, $k$ is the magnitude of the scattering vector, $V$ is the
volume of the system, $\rho _i$ the number density of species $i.$ Further, $%
F_i(k)=F_i^0B_i\left( k\right) $ denotes the scattering amplitude (or form
factor) of species $i$, with $F_i^0$ being the scattering amplitude at zero
angle and $B_i\left( k\right) $ the (angular averaged) intraparticle
interference factor. The components of the column vector ${\bf f}\left(
k\right) $ and of its transpose, the row vector ${\bf f}^T\left( k\right) 
$, are the form factors $F_i\left( k\right) $ weighted by means of the
corresponding densities. More precisely, we define

\begin{equation}
f_i(k)=\rho _i^{1/2}F_i(k)  \label{f2}
\end{equation}

\noindent In order to get more compact formulas, often our notation slightly
departs from Vrij's one \cite{Vrij79}. Specifically, here this author uses $%
f_i$ to denote our $F_i^0$.

Finally, ${\bf S}\left( k\right) $ is a symmetric matrix whose elements
are the Ashcroft-Langreth partial structure factors $S_{ij}\left( k\right) $%
, defined by

\begin{equation}
S_{ij}\left( k\right) =\delta _{ij}+H_{ij}\left( k\right) =\delta
_{ij}+\left( \rho _i\rho _j\right) ^{1/2}\widetilde{h}_{ij}\left( k\right) ,
\label{f3}
\end{equation}

\noindent or, more concisely,

\begin{equation}
{\bf S}\left( k\right) ={\bf I}+{\bf H}\left( k\right) ,  \label{f4}
\end{equation}

\noindent with ${\bf I}$ being the unit matrix of order $p$ ($\delta
 _{ij} $ = Kronecker delta)$.$ Here, $\widetilde{h}_{ij}\left( k\right) $ is
the three-dimensional Fourier transform of the total correlation function, $
h_{ij}\left( r\right) \equiv g_{ij}\left( r\right) -1,$ and $g_{ij}\left(
r\right) $ is the radial distribution function between two particles of
species $i$ and $j$ at a distance $r.$

In the short-wavelength limit $\left( k\rightarrow \infty \right) ,$ when
the particles scatter the incident radiation independently, one has ${\bf
S}\left( k\rightarrow \infty \right) ={\bf I}$ and therefore

\begin{equation}
R\left( k\rightarrow \infty \right) =\sum_{i=1}^p\rho _iF_i^2\left( k\right)
={\bf f}^T\left( k\right) {\bf f}\left( k\right) =\left| {\bf f}%
\left( k\right) \right| ^2  \label{f5}
\end{equation}

\noindent where $\left| {\bf f}\left( k\right) \right| $ denotes the
magnitude of ${\bf f}\left( k\right) $. The normalized scattering
intensity is also called the {\it effective structure factor} or the
{\it measured average structure factor}

\begin{equation}
S^M\left( k\right) \equiv \frac{I\left( k\right) }{I\left( k\rightarrow
\infty \right) }=\frac{{\bf f}^T\left( k\right) {\bf S}\left( k\right) 
{\bf f}\left( k\right) }{\left| {\bf f}\left( k\right) \right| ^2}=1+%
\frac{{\bf f}^T\left( k\right) {\bf H}\left( k\right) {\bf f}\left(
k\right) }{\left| {\bf f}\left( k\right) \right| ^2}  \label{f6}
\end{equation}

\bigskip

\subsection{Direct correlation functions}

A second possible representation of the scattering functions can be obtained
in terms of the three-dimensional Fourier transforms $\widetilde{c}%
_{ij}\left( k\right) $ of the direct correlation functions $c_{ij}\left(
r\right) .$ These functions are defined by the Ornstein-Zernike (OZ)
integral equations of the liquid state theory, which, for systems with
spherically symmetric interactions, read as

\begin{equation}
h_{ij}\left( r\right) =c_{ij}\left( r\right) +\sum_{m=1}^p\rho _m\int d%
{\bf r}^{\prime }\ c_{im}\left( r^{\prime }\right) h_{mj}\left( |{\bf
r-r}^{\prime }| \right)  \label{f7}
\end{equation}

Note that these equations can be solved only if they are coupled with a
second relationship between $c_{ij}\left( r\right) $ and $h_{ij}\left(
r\right) .$ Such a ``closure'' consists of the exact formula

\begin{equation}
h_{ij}\left( r\right) =\exp \left[ -\beta \phi _{ij}\left( r\right) +\gamma
_{ij}\left( r\right) +B_{ij}\left( r\right) \right] -1,  \label{f8}
\end{equation}

\noindent plus an approximation to the ``bridge'' function $B_{ij}\left(
r\right) ,$ which is a complicated functional of $h_{ij}\left( r\right) $
and higher order correlation functions \cite{Hansen86} ($\ \phi _{ij}\left(
r\right) $ is the interparticle potential, $\beta \equiv 1/\left(
k_BT\right) ,$ $k_B$ is Boltzmann constant and $T$ the absolute temperature; 
$\gamma _{ij}\left( r\right) \equiv h_{ij}\left( r\right) -c_{ij}\left(
r\right) \ $).

By Fourier transforming the OZ convolution equations, these can be written
in $k$-space as

\begin{equation}
\left[ {\bf I}+{\bf H}\left( k\right) \right] \left[ {\bf I}-%
{\bf C}\left( k\right) \right] ={\bf I,}  \label{f9}
\end{equation}

\noindent with $C_{ij}\left( k\right) \equiv \left( \rho _i\rho _j\right)
^{1/2}\widetilde{c}_{ij}\left( k\right) .$ Since ${\bf S}\left( k\right) =%
{\bf I}+{\bf H}\left( k\right) ,$ we also get

\begin{equation}
{\bf S}\left( k\right) =\left[ {\bf I}-{\bf C}\left( k\right)
\right] ^{-1},  \label{f10}
\end{equation}

\noindent or, equivalently,

\begin{equation}
S_{ij}\left( k\right) =\frac{\ \left| {\bf I}-{\bf C}\left( k\right)
\right| ^{ji}}{\left| {\bf I}-{\bf C}\left( k\right) \right| },
\label{f11}
\end{equation}

\noindent where $\left| {\bf I}-{\bf C}\left( k\right) \right| $ is
the $p\times p$ determinant of the matrix ${\bf I}-{\bf C}\left(
k\right) $ and $\left| {\bf I}-{\bf C}\left( k\right) \right| ^{ji}$
is the cofactor of its $\left( j,i\right) $th element.

By substituting eq. ($\ref{f10}$) into eq. ($\ref{f1}$), we then obtain the
scattering intensity in terms of the $\widetilde{c}_{ij}\left( k\right) $,
i.e.,

\begin{equation}
R(k)={\bf f}^T\left( k\right) \left[ {\bf I}-{\bf C}\left( k\right)
\right] ^{-1}{\bf f}\left( k\right) ,  \label{f12}
\end{equation}

\noindent while the average structure factor is always given by $%
S^M(k)=R(k)/\left| {\bf f}\left( k\right) \right| ^2.$

For theoretical investigations all these formulas may be more useful than
those in terms of the $\widetilde{h}_{ij}\left( k\right) .$ In fact, in the
cases in which the OZ equations have been analytically solved (for some
particular interparticle potentials and with appropriate ``closures'') the $%
c_{ij}\left( r\right) $ have, in general, rather simple expressions, whereas
the $h_{ij}\left( r\right) $ do not admit a simple analytic representation
and are usually evaluated by numerical inverse Fourier transform of $%
\widetilde{h}_{ij}\left( k\right) $ (in $k$-space, the $\widetilde{h}%
_{ij}\left( k\right) $ are again much more involved than the $\widetilde{c}%
_{ij}\left( k\right) \ $).

\subsection{Baxter factor correlation functions}

Vrij \cite{Vrij79} proposed a third representation of the scattering
functions, in terms of the so-called Baxter factor correlation functions $%
q_{ij}\left( r\right) .$ Baxter \cite{Baxter70} showed, for hard sphere
fluids, that the OZ equations can be transformed in an equivalent, but often
easier to solve, form, by introducing a Wiener-Hopf factorization of the
matrix ${\bf I}-{\bf C}\left( k\right) .$ Later on Hiroike \cite
{Hiroike79} extended Baxter's work to disordered fluids with any kind of
spherically symmetric potentials and obtained generalized Baxter equations
without using the Wiener-Hopf factorization.

Noting that ${\bf I}-{\bf C}(k)$ is a symmetric matrix and an even
function of $k,$ Baxter \cite{Baxter70} suggested the following factorization

\begin{equation}
{\bf I}-{\bf C}(k)=\widehat{{\bf Q}}^T\left( -k\right) \widehat{%
{\bf Q}}\left( k\right)  \label{f13}
\end{equation}

\noindent where the elements of $~\widehat{{\bf Q}}\left( k\right) $ are
of the form 
\begin{equation}
\widehat{{\bf Q}}\left( k\right) ={\bf I}-\widetilde{{\bf Q}}\left(
k\right) ,\qquad \mbox{with \qquad }\widetilde{{\bf Q}}\left( k\right)
=\int_{-\infty }^{+\infty }dr\ e^{ikr}{\bf Q}\left( r\right)  \label{f14}
\end{equation}

\noindent and

\begin{equation}
Q_{ij}(r)=2\pi \left( \rho _i\rho _j\right) ^{1/2}q_{ij}(r)  \label{f15}
\end{equation}

\medskip \noindent ($\ \widehat{{\bf Q}}^T$ is the transpose of $\widehat{%
{\bf Q}}\ $). In general, in the analytically solvable cases, the factor
correlation functions $q_{ij}(r)$ and $\widehat{Q}_{ij}(k)$ have a even
simpler mathematical form than $c_{ij}(r)$ and $C_{ij}(k),$ respectively.

Substituting eq. (\ref{f13}) into eq. (\ref{f10}) yields

\begin{equation}
{\bf S}(k)=\widehat{{\bf Q}}^{-1}\left( k\right) \left[ \widehat{%
{\bf Q}}^{-1}\left( -k\right) \right] ^T  \label{f16}
\end{equation}

\noindent and, since

\begin{equation}
\widehat{Q}_{ij}^{-1}\left( k\right) =\frac{\ \left| \widehat{{\bf Q}}%
\left( k\right) \right| ^{ji}}{\left| \widehat{\bf{Q}}\left( k\right)
\right| },  \label{f17}
\end{equation}

\noindent then

\begin{equation}
S_{ij}(k)=\sum_m\widehat{Q}_{im}^{-1}\left( k\right) \widehat{Q}%
_{jm}^{-1}\left( -k\right) =\frac 1{D(k)}\sum_m\left| \widehat{\bf{Q}}%
\left( k\right) \right| ^{mi}\left| \widehat{\bf{Q}}\left( -k\right)
\right| ^{mj}  \label{f18}
\end{equation}

\noindent with

\begin{equation}
D(k)\equiv \left| \widehat{\bf{Q}}\left( k\right) \right| \left| 
\widehat{\bf{Q}}\left( -k\right) \right| =\left| {\bf I}-{\bf C}%
\left( k\right) \right|  \label{f19}
\end{equation}

\medskip

For the scattering intensity, plugging eq. (\ref{f16}) into eq. (\ref{f1})
and assuming that $\bf{f}\left( k\right) ={\bf f}\left( -k\right) $
yields the simple expression

\begin{equation}
R(k)=\bf{s}^T\left( k\right) {\bf s}\left( -k\right) =\frac 1{D(k)}%
\bf{L}^T\left( k\right) {\bf L}\left( -k\right) ,  \label{f20}
\end{equation}

\noindent where we have defined 
\begin{equation}
\bf{s}\left( k\right) \equiv \left[ \widehat{{\bf Q}}^{-1}\left(
k\right) \right] ^T\bf{f}\left( k\right) \equiv \frac{{\bf L}\left(
k\right) }{\left| \widehat{\bf{Q}}\left( k\right) \right| }  \label{f21}
\end{equation}

\medskip

\noindent The hypothesis $\bf{f}\left( k\right) ={\bf f}\left(
-k\right) ,$ which is equivalent to \bigskip $F_i\left( k\right) =F_i\left(
-k\right) ,$ is really correct for the physical systems we are concerned
with in this paper. In fact, for homogeneous spheres with radius $\sigma
_i/2 $, it results that

\begin{equation}
F_i\left( k\right) \propto V_i\Phi _1\left( k\sigma _i/2\right) ,
\label{f22}
\end{equation}

\noindent where $V_i=\left( \pi /6\right) \sigma _i^3$ is the volume of a
particle of species $i$ and $\Phi _1\left( x\right) $ is an even function
which may be expressed in terms of the first-order spherical Bessel function 
$j_1\left( x\right) $ as

\begin{equation}
\Phi _1\left( x\right) \equiv 3j_1\left( x\right) /x=3\left( \sin x-x\cos
x\right) /x^3  \label{f23}
\end{equation}

Eq.s (\ref{f16}) and (\ref{f20}) give the partial structure factors and the
scattering intensity in terms of the Baxter functions $\widehat{Q}_{ij}(k).$

\section{The Analytically tractable case of dyadic matrices}

In view of producing theoretical expressions for the scattering functions,
we have presented three possible routes, based upon different correlation
functions, i.e., $H_{ij}\left( k\right) ,$ $C_{ij}\left( k\right) $ and $%
\widehat{Q}_{ij}(k),$ respectively.

In Baxter's route, followed in this paper, the crucial point for the
analytic evaluation of the scattering functions is the possibility of
getting a simple and closed expression for the inverse $\widehat{\bf{Q}}%
^{-1}\left( k\right) $ of the matrix $\widehat{\bf{Q}}\left( k\right) .$
We want to emphasize that such a task can easily be accomplished if $%
\widehat{Q}_{ij}(k)$ has a particularly convenient structure of the
following general form

\begin{equation}
\widehat{Q}_{ij}(k)=\delta _{ij}+\sum_{\mu =1}^na_i^{(\mu )}b_j^{(\mu
)}\equiv \delta _{ij}+\widehat{W}_{ij}\left( k\right)  \label{f24}
\end{equation}

\noindent Clearly, $\widehat{W}_{ij}\left( k\right) $ is simply related to
the unidimensional Fourier transform $\widetilde{q}_{ij}\left( k\right) $ of 
$q_{ij}\left( r\right) ,$ i.e., $\widehat{W}_{ij}\left( k\right) =-2\pi
\left( \rho _i\rho _j\right) ^{1/2}\widetilde{q}_{ij}\left( k\right) .$

It is now convenient to recall that a second-rank tensor $T_{ij}=a_ib_j$
formed by the outer product of two vectors is sometimes represented by a
symbol called a {\it dyad}, $\bf{ab},$ and a linear combination of
dyads $\sum_{\mu} \lambda _{\mu}\bf{a}^{\left( \mu \right)} {\bf b}^{\left(
\mu \right) }$ is called a {\it dyadic} \cite{Mathews65}. In addition, as
we propose here, a dyadic formed by a sum of $n$ dyads might be called an 
{\it n-dyadic.}

Using this terminology, we could say that eq. (\ref{f24}) is equivalent to
require that $\widetilde{q}_{ij}\left( k\right) $ is a generic $n$-dyadic
matrix (of order $p$).

In the case of the PY solution for neutral HS mixtures (including the
polydisperse limit), $\widetilde{q}_{ij}\left( k\right) $ is a $2$-dyadic
matrix \cite{Vrij79}, \cite{Stell79}. In the MSA solution for charged HS \cite
{Blum77}, as we shall better see in Section V, $\widetilde{q}_{ij}\left(
k\right) $ becomes a $3$-dyadic.

Both Vrij \cite{Vrij79} and Blum with coworkers \cite{Stell79}, \cite{Blum77}
investigated some properties of these matrices, independently and from
complementary points of view. On the one hand, Vrij emphasized the role of
the rank of $\widehat{\bf{Q}}\left( k\right) $ and developed an original
method to derive $\widehat{\bf{Q}}^{-1}\left( k\right) .$ On the other
hand, Blum and coworkers gave more directly closed expressions for $\widehat{%
\bf{Q}}^{-1}\left( k\right) $ for both 2- and 3-dyadics (on passing, we
note that none of these authors uses our terminology; for instance, Blum and
H\o ye \cite{Blum77} refer to these matrices as Jacobi matrices, whose
inverse can be found for any size of the matrix).

In order to join and generalize the two above-mentioned approaches, in this
Section we shall present some {\it general} properties of $n$-dyadic
matrices, which hold true for any $n$ (the relevant proofs are outlined in
Appendix A).

Let us start from the fact that for any matrix of the form

\begin{equation}
M_{ij}=\delta _{ij}+\widehat{W}_{ij},  \label{f25}
\end{equation}

\noindent where $\widehat{W}_{ij}$ is a generic $p\times p$ matrix (i.e.,
not necessarily a dyadic), the determinant $\left| \bf{M}\right| $ can
be expressed by the following expansion in terms of principal minors 
\begin{eqnarray}
\left| \bf{M}\right| &=&\left| {\bf I+\widehat{W}}\right|
=1+\sum_{i=1}^p\widehat{W}_{ii}+\frac 1{2!}\sum_{i,j=1}^p\left| 
\begin{array}{ll}
\widehat{W}_{ii} & \widehat{W}_{ij} \\ 
\widehat{W}_{ji} & \widehat{W}_{jj}
\end{array}
\right|  \nonumber \\
&&  \label{f26} \\
&&+\frac 1{3!}\sum_{i,j,k=1}^p\left| 
\begin{array}{lll}
\widehat{W}_{ii} & \widehat{W}_{ij} & \widehat{W}_{ik} \\ 
\widehat{W}_{ji} & \widehat{W}_{jj} & \widehat{W}_{jk} \\ 
\widehat{W}_{ki} & \widehat{W}_{kj} & \widehat{W}_{kk}
\end{array}
\right| +\cdots  \nonumber
\end{eqnarray}

For a generic $\widehat{W}_{ij}$ matrix, this expansion clearly stops at the 
$p$-th term. However, it is easy to show that, if $\widehat{W}_{ij}$ is an $n
$-dyadic (with $n<p$), then any its minor having order $m>n$ vanishes! Using
this along with some manipulations, we find our first basic result on dyadic
matrices: {\it any }$p\times p$ matrix of the form

\begin{equation}
{\bf M}= {\bf I}+\sum_{\mu =1}^n {\bf a}^{(\mu )}{\bf b}^{(\mu )}
\label{f27}
\end{equation}

\noindent {\it always has rank } $n,$ {\it irrespective of }$p.$ 
{\it Consequently, the expansion (\ref{f26}) of its determinant breaks
off after the }$n${\it th term and }$\left| \bf{M}\right| $,{\it 
which is of order }$p$,{\it turns out to be equal to a determinant }$D_M$
{\it of order }$n<p,${\it as expressed synthetically by}

\begin{equation}
\left| \bf{M}\right| =D_M\equiv \det \left( \delta _{\mu \nu }+{\bf a}%
^{(\mu )}\cdot \bf{b}^{(\nu )}\right) _n  \label{f28}
\end{equation}

\noindent where $\det \left( ...\right) _n$ means the determinant of a
matrix of order $n$, while the dot denotes the usual scalar product of
vectors

\begin{equation}
\bf{a}^{(\mu )}\cdot {\bf b}^{(\nu )}=\sum_{m=1}^pa_m^{(\mu
)}b_m^{(\nu )}  \label{f29}
\end{equation}

It is to be noted that the use of expansion (\ref{f26}) with $2$-dyadics is
originally due to Vrij \cite{Vrij79}; the extension to $n>2$ and eq. (\ref
{f28}) are new.

Let us now consider the cofactors $\left| \bf{M}\right| ^{ji}$ which
will be needed in the calculation of the inverse matrix $\widehat{\bf{Q}}%
^{-1}\left( k\right) $ and, consequently, of the scattering intensity, as we
shall discuss in Section V. In the same Appendix A it is also shown that for
a matrix of the type (\ref{f25}) with a {\it generic} $\widehat{W}_{ij}$
the cofactors can be written as follows

\begin{equation}
\left| \bf{M}\right| ^{ji}=\left| {\bf M}\right| \delta _{ij}-%
\widehat{U}_{ij}  \label{f30}
\end{equation}

\noindent where the following expansion for $\widehat{U}_{ij}$ holds true

\begin{equation}
\widehat{U}_{ij}=\widehat{W}_{ij}+\sum_{k=1}^p\left| 
\begin{array}{ll}
\widehat{W}_{ij} & \widehat{W}_{ik} \\ 
\widehat{W}_{kj} & \widehat{W}_{kk}
\end{array}
\right| +\frac 1{2!}\sum_{k,l=1}^p\left| 
\begin{array}{lll}
\widehat{W}_{ij} & \widehat{W}_{ik} & \widehat{W}_{il} \\ 
\widehat{W}_{kj} & \widehat{W}_{kk} & \widehat{W}_{kl} \\ 
\widehat{W}_{lj} & \widehat{W}_{lk} & \widehat{W}_{ll}
\end{array}
\right| +\cdots  \label{f31}
\end{equation}

Again for $n$-dyadic matrices this expansion breaks off at the level $n<p$,
as it is straightforward to check. Along the same lines as before, we find
that $\widehat{U}_{ij}$ can be written as follows

\begin{equation}
\widehat{U}_{ij}=\sum_{\mu =1}^na_i^{(\mu )}\widehat{D}_j^{(\mu )},
\label{f32}
\end{equation}

\noindent where the $\widehat{D}_j^{(\mu )}$ are $n\times n$ determinants
which can be obtained from the determinant $D_M$ upon replacing the $\mu $%
-th {\it row} with a row formed by $b_j^{(1)},...,b_j^{(n)}.$

From $M_{ij}^{-1}=\left| \bf{M}\right| ^{ji}/\left| {\bf M}\right| ,$
it is now immediate to get the inverse matrix $\bf{M}^{-1}$

\begin{equation}
M_{ij}^{-1}=\delta _{ij}-\sum_{\mu =1}^na_i^{(\mu )}\frac{\widehat{D}%
_j^{(\mu )}}{D_M},  \label{f33}
\end{equation}

\noindent with

\begin{equation}
\widehat{D}_j^{(\mu )}\equiv \det \left( \ \left[ 1-\delta _{\alpha \mu
}\right] \left[ \delta _{\alpha \beta }+\bf{a}^{(\alpha )}\cdot {\bf b%
}^{(\beta )}\right] +\delta _{\alpha \mu }\ b_j^{(\beta )}\ \right) _n
\label{f34}
\end{equation}

This general formula constitutes our second result on dyadic matrices,
yielding a new simple and elegant recipe which holds true for any $n$ value: 
{\it the inverse of any }$p\times p$ {\it matrix }${\bf M}={\bf I}
+\sum_{\mu =1}^n\bf{a}^{(\mu )} {\bf b}^{(\mu )}$ {\it has
elements given by eq. (\ref{f33}) as linear functions of simple determinants 
}$\widehat{D}_j^{(\mu )}${\it of order }$n$.{\it The determinant }$%
\widehat{D}_j^{(\mu )}$ {\it is obtained from }$D_M$,{\it defined in
eq. (\ref{f28}), by using a Cramer-like rule, i.e., by replacing the }$\mu $%
{\it -th row with a row formed by} $b_j^{(1)},...,b_j^{(n)}$ (of course,
all rows can be interchanged with the corresponding columns, without
altering the value of the determinant).

Due to the symmetric role played by the $a$'s and the $b$'s in eq. (\ref{f28}%
), it is clear that one could derive an alternative expression to (\ref{f32}%
) with the role of the the $a$'s and the $b$'s interchanged.

Note that our formulas for $\left| \bf{M}\right| ^{ji}$ and $M_{ij}^{-1}$
are fully new with respect to Vrij's work \cite{Vrij79}. Eq. (\ref{f33})
agrees with the expressions of $M_{ij}^{-1}$ given by Blum and coworkers 
\cite{Stell79}, \cite{Blum77} for $n=2$ and $n=3$; at the same time, our
generalization offers a systematic way to calculate $\bf{M}^{-1}$ for
any $n.$

We shall be using all these results in Section V.

\section{Charged Hard Spheres and MSA solution}

The {\it primitive model }for ionic fluids consists of a $p$-component
electroneutral mixture of charged hard spheres embedded in a continuum of
dielectric constant $\varepsilon $ (which may represent a possible solvent)$%
. $ The species $i$ has diameter $\sigma _i,$ number density $\rho _i$ and
electric charge $z_ie$ ($e$ is the elementary charge). The interparticle
potential is the hard sphere Coulombic one, i.e.,

\begin{equation}
\phi _{ij}\left( r\right) =\left\{ 
\begin{array}{l}
+\infty \qquad \qquad \mbox{for\qquad }r<\sigma _{ij}\equiv \frac 12\left(
\sigma _i+\sigma _j\right) \\ 
\\ 
e^2z_iz_j/(\varepsilon r)\qquad \qquad \mbox{for\qquad }r>\sigma _{ij}
\end{array}
\right.  \label{f35}
\end{equation}

\noindent and the electroneutrality condition reads

\begin{equation}
\sum_m\rho _mz_m=0  \label{f36}
\end{equation}

In $r$-space the Baxter form of the OZ equations which can be derived from
eq. (\ref{f13}) is \cite{Baxter70}, \cite{Hiroike79}

\begin{equation}
\left\{ 
\begin{array}{l}
rc_{ij}\left( |r|\right) =-q_{ij}^{\ \prime }(r)+2\pi \sum_m\rho
_m\int_{\lambda _{mi}}^\infty dt\ q_{mi}\left( t\right) q_{mj}^{\ \prime
}\left( r+t\right) ,\qquad r>\lambda _{ij}\equiv \frac 12\left( \sigma
_i-\sigma _j\right) \\ 
\\ 
rh_{ij}\left( |r|\right) =-q_{ij}^{\ \prime }(r)+2\pi \sum_m\rho
_m\int_{\lambda _{im}}^\infty dt\ q_{im}\left( t\right) \left( r-t\right)
h_{mj}\left( |r-t|\right) ,\qquad r>\lambda _{ij}
\end{array}
\right.  \label{f37}
\end{equation}

\bigskip

\noindent where the prime denotes differentiation with respect to $r$. Note
that, if $\lambda _{ij}<0,$ then $r$ may assume negative values as well,
although the correlation functions $c_{ij}(r)$ and $h_{ij}\left( r\right) $
are determined, of course, only for positive distances.

The Mean Spherical Approximation, used to solve these equations, consists of
adding to the exact hard core condition

\begin{equation}
h_{ij}\left( r\right) =-1\qquad \qquad \mbox{for\qquad }r<\sigma _{ij},
\label{f38}
\end{equation}

\bigskip \noindent the approximate relationship

\begin{equation}
c_{ij}\left( r\right) =-\beta \phi _{ij}(r)\qquad \qquad \mbox{for\qquad }%
r>\sigma _{ij},  \label{f39}
\end{equation}

\medskip

\noindent which is asymptotically correct for $r\rightarrow \infty .$ For
uncharged HS potentials, the MSA coincides with the PY approximation: $%
c_{ij}\left( r\right) =\left\{ \exp \left[ -\beta \phi _{ij}(r)\right]
-1\right\} \left[ 1+h_{ij}\left( r\right) -c_{ij}\left( r\right) \right] .$
The MSA, eq. (\ref{f39}), may also be regarded as the following
approximation to the bridge functions

\begin{equation}
B_{ij}\left( r\right) =\ln \left[ 1+\gamma _{ij}(r)-\beta \phi
_{ij}(r)\right] -\left[ \gamma _{ij}(r)-\beta \phi _{ij}(r)\right] \qquad
\qquad \mbox{for\qquad }r>\sigma _{ij}  \label{f40}
\end{equation}

\medskip

The solution of the MSA integral equations for the most general case of ions
with arbitrary charges and diameters was found by Blum \cite{Blum75}, \cite
{Blum77}, \cite{Blum80A}, \cite{Blum80B}$.$ It is worth noting that Blum
employed a form of the Baxter equations which is slightly different from
eq.s (\ref{f37}). In this paper it proves convenient to follow the original
 form proposed by Baxter \cite{Baxter70} and Hiroike \cite{Hiroike79}, eq.s (%
 \ref{f37}), and used by Vrij \cite{Vrij79}. The differences from Blum's
 formulas of ref. \cite{Blum77} are, however, small  (a detailed comparison
 is given in Appendix B).
Our modified form of the MSA solution reads

\begin{equation}
q_{ij}\left( r\right) =\left\{ 
\begin{array}{l}
0,\qquad \qquad \qquad r<\lambda _{ij} \\ 
\\ 
\frac 12\left( r-\sigma _{ij}\right) ^2q_i^{\ \prime \prime }+\left(
r-\sigma _{ij}\right) q_{ij}^{\ \prime }-A_iz_j,\qquad \lambda
_{ij}<r<\sigma _{ij} \\ 
\\ 
-A_iz_j,\qquad \qquad r>\sigma _{ij}
\end{array}
\right.  \label{f41}
\end{equation}

\medskip

The coefficients of the factor correlation functions are given by

\begin{equation}
q_{ij}^{\ \prime }=\frac 1\Delta \left( \sigma _{ij}+\frac{3\xi _2}{2\Delta }%
\sigma _i\sigma _j\right) -\frac{\Gamma ^2}{\ell _B}A_iA_j  \label{f42}
\end{equation}

\begin{equation}
q_i^{\ \prime \prime }=\frac 1\Delta \left( 1+\frac{3\xi _2}\Delta \sigma
_i\right) + A_iP_z  \label{f43}
\end{equation}

\begin{equation}
 A_i=\frac{\ell _B}\Gamma \ \frac{z_i-\frac 12\sigma _i^2P_z}{1+\Gamma \sigma
 _i}  \label{f44}
\end{equation}

\noindent with

\begin{equation}
\xi _n=\frac \pi 6\sum_m\rho _m\sigma _m^n\qquad (n=2,3)  \label{f45}
\end{equation}

\begin{equation}
\Delta =1-\xi _3  \label{f46}
\end{equation}

\begin{equation}
\ell _B=\beta e^2/\varepsilon \qquad (\mbox{Bjerrum length})  \label{f47}
\end{equation}

\begin{equation}
P_z=\frac 1\Omega \sum_m\frac{\rho _m\sigma _mz_m}{1+\Gamma \sigma _m}
\label{f48}
\end{equation}

\begin{equation}
\Omega =1-\xi _3+\frac \pi 2\sum_m\frac{\rho _m\sigma _m^3}{1+\Gamma \sigma
_m}  \label{f49}
\end{equation}

\medskip

Finally, the value of the parameter $\Gamma $ can be determined by solving
numerically the following consistency equation

\begin{equation}
\left( 2\Gamma \right) ^2=4\pi \ell _B\sum_m\rho _m\left( \frac{z_m-\frac{1}
{2}\sigma _m^2P_z}{1+\Gamma \sigma _m}\right) ^2,  \label{f50}
\end{equation}

\noindent which is equivalent to the condition

\begin{equation}
\sum_m\rho _mA_m^2=\ell _B/\pi  \label{f51}
\end{equation}

In the limit of point ions (all $\sigma _m\rightarrow 0$), $2\Gamma $ tends
to the Debye inverse shielding length $\kappa _{D\mbox{ }}$ of the classical
Debye-H\"{u}ckel theory for electrolyte solutions, while for finite size
ions $\left( 2\Gamma \right) ^{-1}$ is always larger than $\kappa _{D\mbox{ }%
}$:

\begin{equation}
2\Gamma \leq \kappa _{D\mbox{ }}\equiv \left( 4\pi \ell _B\sum_m\rho
_mz_m^2\right) ^{1/2}  \label{f52}
\end{equation}

There is a striking similarity of the MSA to the Debye-H\"{u}ckel theory,
with $2\Gamma $ taking the place of $\kappa _D$ ($2\Gamma $ is the correct
screening parameter for finite size ions).

\noindent It is to be noted that the degree of the algebraic equation (\ref
{f50}) for $\Gamma $ increases with the number $p$ of components in the
mixture. Moreover, among all its solutions, only one is physically
acceptable: the one which is positive and tendes to $\kappa _{D\mbox{ }}$
from below in the infinite dilution limit.

\section{Application of Dyadic Properties to Charged Hard Spheres}

\subsection{The inverse matrix $\widehat{\bf{Q}}^{-1}\left( k\right) $}

The starting point is the evaluation of the unidimensional Fourier
transform, $\widetilde{q}_{ij}\left( k\right) $, of the MSA solution for $%
q_{ij}\left( r\right) $. The resulting $\widehat{Q}_{ij}\left( k\right) $
can be written as

\begin{equation}
\widehat{Q}_{ij}\left( k\right) =\delta _{ij}+\widehat{W}_{ij}\left(
k\right) =\delta _{ij}+\left( \rho _i\rho _j\right) ^{1/2}W_{ij}\left(
k\right) ,  \label{f53}
\end{equation}

\noindent with

\begin{eqnarray}
W_{ij}\left( k\right) &=&-2\pi \widetilde{q}_{ij}\left( k\right)
=e^{ik\sigma _i/2}\left[ \alpha _j\left( k\right) +\sigma _i\beta _j\left(
k\right) +A_i\gamma _j\left( k\right) \right]  \nonumber \\
&&  \label{f54} \\
&=&Z_{ij}\left( k\right) +e^{iX_i}A_i\gamma _j\left( k\right)  \nonumber
\end{eqnarray}

\noindent 
where $i$ \ -\ when it is not a subscript\ -\ is the imaginary
unit and

\begin{equation}
X_m=k\sigma _m/2  \label{f55}
\end{equation}

\begin{equation}
\alpha _j\left( k\right) =\frac{4\pi }\Delta \frac 1{k^3}\left( \sin
X_j-X_j\cos X_j\right) =\frac \pi 6\sigma _j^3\frac 1\Delta \Phi _1\left(
X_j\right)  \label{f56}
\end{equation}

\begin{equation}
\beta _j\left( k\right) =\beta _{0,\ j}\left( k\right) +\beta _{1,\ j}\left(
k\right)   \label{f57}
\end{equation}

\begin{eqnarray}
\beta _{0,\ j}\left( k\right)  &=&\frac{2\pi }\Delta \frac{\sigma _j}k\sin
X_j\equiv \frac \pi 6\sigma _j^2\frac 1\Delta \Phi _0\left( X_j\right) 
\label{f58} \\
&&  \nonumber \\
\beta _{1,\ j}\left( k\right)  &=&\left( \frac{3\xi _2}\Delta -\frac
12ik\right) \alpha _j\left( k\right)   \label{f59}
\end{eqnarray}

\begin{equation}
\gamma _j\left( k\right) =\gamma _{0,\ j}\left( k\right) +\gamma _{1,\
j}\left( k\right) +\gamma _{2,\ j}\left( k\right)   \label{f60}
\end{equation}

\begin{eqnarray}
\gamma _{0,\ j}\left( k\right)  &=&\frac{\Gamma ^2A_j}{\ell _B}\ \frac{4\pi i%
}{k^2}\sin X_j  \label{f61} \\
&&  \nonumber \\
\gamma _{1,\ j}\left( k\right)  &=&\frac{\Gamma A_j}{\ell _B}\ \frac{2\pi i}%
ke^{-iX_j}  \label{f62} \\
&&  \nonumber \\
 \gamma _{2,\ j}\left( k\right)  &=&P_z\ \Delta \left[ \alpha _j\left(
 k\right) +\frac{2i}k\beta _{0,\ j}\left( k\right) \right]   \label{f63}
\end{eqnarray}

\medskip

Here, the function $\Phi _1\left( x\right) =3j_1(x)/x$ is the same as in eq.
(\ref{f23}) and we can also write $\Phi _0\left( x\right) $ in terms of
Bessel functions: $\Phi _0\left( x\right) =3j_0(x)=3\sin x/x$.

Note that $\alpha _j\left( k\right) $ and $\beta _{0,\ j}\left( k\right) $
are even functions, unlike $\beta _j\left( -k\right) =\beta _j^{*}\left(
k\right) $ and $\gamma _j\left( -k\right) =\gamma _j^{*}\left( k\right) $
(the asterisk denotes complex conjugation). Hence, we have $W_{ij}\left(
-k\right) =W_{ij}^{*}\left( k\right) $ and $\widehat{Q}_{ij}\left( -k\right)
=\widehat{Q}_{ij}^{*}\left( k\right) .$

Moreover, $Z_{ij}\left( k\right) $ is the pure hard sphere term considered
by Vrij, while $e^{iX_i}A_i\gamma _j\left( k\right) $ is the electrostatic
contribution ($\alpha _j$ and $\beta _j$ correspond to $M_j$ and $N_j$ of
ref. \cite{Vrij79}, respectively).

In the limit $\left\{ \mbox{all charges }z_m\right\} \rightarrow 0,$ $%
A_i,P_z $ and $\gamma _j(k)$ vanish simultaneously. In the long wavelength
limit ($k\rightarrow 0$), $Z_{ij}(k)$ tends to a finite value, whereas $%
\gamma _j(k)$ diverges as $k^{-1}.$ Such a singularity of $\widetilde{q}%
_{ij}\left( k\right) $ implies a $k^{-2}$ divergence of $\widetilde{c}%
_{ij}\left( k\right) $; however, all this does not influence the structure
factors, which remain finite at the origin.

The crucial point is now that, according to eq. (\ref{f54}), the MSA
expression of $W_{ij}\left( k\right) $ for charged HS mixtures is a $3$%
-dyadic. As we discussed in Sec. II, this peculiar form of $\widehat{Q}%
_{ij}\left( k\right) $ allows to write down a simple and compact expression
for its $p\times p$ determinant $\left| \widehat{\bf{Q}}\left( k\right)
\right| =\left| \bf{I-}\widetilde{{\bf Q}}\left( k\right) \right| $.
Using the result (\ref{f28}) we get immediately, for any $p$,

\begin{equation}
\left| 
\begin{array}{lll}
\widehat{Q}_{11}\left( k\right) & \cdots & \widehat{Q}_{1M}\left( k\right)
\\ 
\vdots & \ddots & \vdots \\ 
\widehat{Q}_{M1}\left( k\right) & \cdots & \widehat{Q}_{MM}\left( k\right)
\end{array}
\right| =D_Q(k)\equiv \left| 
\begin{array}{lll}
1+\ \left\langle \alpha \right\rangle & \qquad \left\langle \beta
\right\rangle & \qquad \left\langle \gamma \right\rangle \\ 
\qquad \left\langle \sigma \alpha \right\rangle & 1+\left\langle \sigma
\beta \right\rangle & \qquad \left\langle \sigma \gamma \right\rangle \\ 
\qquad \left\langle A\alpha \right\rangle & \qquad \left\langle A\beta
\right\rangle & 1+\left\langle A\gamma \right\rangle
\end{array}
\right| ,  \label{f64}
\end{equation}

\noindent where we have defined

\begin{equation}
\left\langle fg\right\rangle \equiv \sum_m\rho _me^{iX_m}f_m(k)g_m(k)
\label{f65}
\end{equation}

\noindent Notice that our shorthand notation is somewhat different from
Vrij's one. Of course, $\left\langle \alpha \right\rangle $, $\left\langle
\beta \right\rangle $, etc. are complex functions of $k$. However, as in eq.
(\ref{f64}), we shall mostly omit the argument, unless when necessary. The
connection with the $a$'s and $b$'s is

\begin{equation}
\begin{array}{ccc}
a_i^{(1)}=\rho _i^{1/2}e^{iX_i}, &  & b_j^{(1)}=\rho _j^{1/2}\alpha _j \\ 
&  &  \\ 
a_i^{(2)}=\rho _i^{1/2}e^{iX_i}\sigma _{i\ }, &  & b_j^{(2)}=\rho
_j^{1/2}\beta _j \\ 
&  &  \\ 
a_i^{(3)}=\rho _i^{1/2}e^{iX_i}A_i\ , &  & b_j^{(3)}=\rho _j^{1/2}\gamma _j
\end{array}
\label{f66}
\end{equation}

\medskip Eq. (\ref{f64}) clearly stresses the fact that, even if the order
of the determinant $\left| \widehat{\bf{Q}}\left( k\right) \right| $ may
become quite large with increasing the number $p$ of components in the
mixture, nevertheless $D_Q(k)$ remains of order $3$ (\mbox{{\it contraction or
reduction effect}}).

From $\left| \widehat{\bf{Q}}\left( k\right) \right| =D_Q(k)$ and eq. (%
\ref{f19}), it follows that

\begin{equation}
D(k)=\left| \bf{I}-{\bf C}\left( k\right) \right|
=D_Q(k)D_Q(-k)=D_Q(k)D_Q^{*}(k)  \label{f67}
\end{equation}

Once again, $D(k)=\left| \bf{I}-{\bf C}\left( k\right) \right| $ is a
determinant of order $p$, whereas $D_Q(k)D_Q(-k)$ is a product of two
determinants of order three. Since $\left| \bf{A}\right| \left| {\bf B%
}\right| =\left| \bf{AB}\right| $ (if ${\bf A}$ and ${\bf B}$ are
any two square matrices of the same order), $D_Q(k)D_Q(-k)$ could be
rewritten as a single determinant of order 3. We can thus conclude that,
under the peculiar assumptions of our model, also the matrix $\bf{I}-%
\bf{C}\left( k\right) $ has always rank 3, no matter how large is the
number $p$ of components.

Similar considerations hold true for the cofactor matrices. From eq.s (\ref
{f30}) and (\ref{f32}) one can easily show, after some manipulations, that

\begin{equation}
\left| \widehat{\bf{Q}}\left( k\right) \right| ^{ji}=\left| \widehat{%
\bf{Q}}\left( k\right) \right| \delta _{ij}-\left( \rho _i\rho _j\right)
^{1/2}e^{iX_i}\left[ D_j^{(1)}(k)+\sigma
_iD_j^{(2)}(k)+A_iD_j^{(3)}(k)\right] ,\qquad (i,j=1,...,p)  \label{f68}
\end{equation}

\noindent where $D_j^{(\mu )}\ (\mu =1,2,3)$ is simply defined as the $%
3\times 3$ determinant derived from $D_Q$ by replacing its $m$th row with
the vector $\left( \alpha _j,\ \beta _j,\ \gamma _j\right) .$

Then, from $\widehat{Q}_{ij}^{-1}\left( k\right) =~\left| \widehat{\bf{Q}%
}\left( k\right) \right| ^{ji}/\left| \widehat{\bf{Q}}\left( k\right)
\right| $, it is now immediate to get the inverse matrix of $\widehat{%
\bf{Q}}\left( k\right) $

\begin{equation}
\widehat{\bf{Q}}^{-1}\left( k\right) ={\bf I-}\widehat{{\bf V}}(k)
\label{f69}
\end{equation}

\noindent with

\begin{equation}
\widehat{V}_{ij}(k)\equiv \left( \rho _i\rho _j\right) ^{1/2}V_{ij}\left(
k\right) \equiv \left( \rho _i\rho _j\right) ^{1/2}e^{iX_i}\left[ \alpha
_j^{\ \prime }(k)+\sigma _i\beta _j^{\ \prime }(k)+A_i\gamma _j^{\ \prime
}(k)\right] ,  \label{f70}
\end{equation}

\begin{equation}
\alpha _j^{\ \prime }(k)\equiv \frac{D_j^{(1)}(k)}{D_Q(k)},\qquad \beta
_j^{\ \prime }(k)\equiv \frac{D_j^{(2)}(k)}{D_Q(k)},\qquad \gamma _j^{\
\prime }(k)\equiv \frac{D_j^{(3)}(k)}{D_Q(k)}  \label{f71}
\end{equation}

\medskip

\noindent Here, the prime does not denote a derivative: we have used this
notation only to emphasize the similarity of the form of $V_{ij}(k)$ to that
of $W_{ij}(k)$ in $\widehat{Q}_{ij}\left( k\right) $.

\subsection{Scattering intensity}

To get the scattering intensity

\begin{equation}
R(k)=\sum_js_j(k)\ s_j(-k),  \label{f73}
\end{equation}

\noindent where

\begin{equation}
s_j(k)=\frac{kL_j\left( k\right) }{kD_Q(k)},  \label{f74}
\end{equation}

\noindent let us first evaluate

\begin{equation}
L_j\left( k\right) =\sum_m\rho _m^{1/2}F_m\left( k\right) \left| \widehat{%
\bf{Q}}\left( k\right) \right| ^{jm}  \label{f75}
\end{equation}

Note that in eq. (\ref{f74}) both the numerator and the denominator have
been multiplied by $k$ to ``heal'' the $k^{-1}$ singularity of $\gamma _j(k)$
at the origin and ensure that in numerical calculations $R(k)$ has a regular
behaviour as $k\rightarrow 0.$

If the expression (\ref{f68}) for $\left| \widehat{\bf{Q}}\left(
k\right) \right| ^{jm}$ is inserted in this equation, then, after some
manipulations, $L_j\left( k\right) $ can be cast in a very convenient form
as a $4\times 4$ determinant, i.e.,

\begin{equation}
L_j\left( k\right) =\rho _j^{1/2}\left| 
\begin{array}{llll}
~F_j & \qquad \alpha _j & \ \qquad \beta _j & \qquad \gamma _j \\ 
\left\langle F\right\rangle & \quad 1+\ \left\langle \alpha \right\rangle & 
\qquad \left\langle \beta \right\rangle & \qquad \left\langle \gamma
\right\rangle \\ 
\left\langle \sigma F\right\rangle & \qquad \left\langle \sigma \alpha
\right\rangle & \quad 1+\left\langle \sigma \beta \right\rangle & \qquad
\left\langle \sigma \gamma \right\rangle \\ 
\left\langle AF\right\rangle & \qquad \left\langle A\alpha \right\rangle & 
\qquad \left\langle A\beta \right\rangle & \quad 1+\left\langle A\gamma
\right\rangle
\end{array}
\right|  \label{f77}
\end{equation}

However, to perform the summation over $j$, it is more convenient to
re-expand this determinant along the first row

\begin{equation}
L_j\left( k\right) =\rho _j^{1/2}\left[ F_j(k)T_1(k)+\alpha
_j(k)T_2(k)+\beta _j(k)T_3(k)+\gamma _j(k)T_4(k)\right] ,  \label{f78}
\end{equation}

\noindent where $T_{\mu} (k)$ $\left( \mu =1,...,4\right) $ is the cofactor of
the $(1,\mu )$th element of the first row. Clearly, $T_1(k)=D_Q(k).$

In this way we obtain

\begin{equation}
R(k)=\left\{ 
\begin{array}{llll}
\left\{ F^2\right\} & +\left\{ F\alpha \right\} C_1^{*} & +\left\{ F\beta
\right\} C_2^{*} & +\left\{ F\gamma \right\} C_3^{*} \\ 
&  &  &  \\ 
+\left\{ \alpha F\right\} C_1 & +\left\{ \alpha ^2\right\} C_1C_1^{*} & 
+\left\{ \alpha \beta ^{*}\right\} C_1C_2^{*} & +\left\{ \alpha \gamma
^{*}\right\} C_1C_3^{*} \\ 
&  &  &  \\ 
+\left\{ \beta F\right\} C_2 & +\left\{ \beta \alpha \right\} C_2C_1^{*} & 
+\left\{ \beta \beta ^{*}\right\} C_2C_2^{*} & +\left\{ \beta \gamma
^{*}\right\} C_2C_3^{*} \\ 
&  &  &  \\ 
+\left\{ \gamma F\right\} C_3 & +\left\{ \gamma \alpha \right\} C_3C_1^{*} & 
+\left\{ \gamma \beta ^{*}\right\} C_3C_2^{*} & +\left\{ \gamma \gamma
^{*}\right\} C_3C_3^{*}
\end{array}
\right.  \label{f79}
\end{equation}

\noindent with

\begin{equation}
\left\{ fg\right\} \equiv \sum_m\rho _mf_m\left( k\right) g_m\left( k\right)
\label{f80}
\end{equation}

\begin{equation}
C_{\mu} (k)\equiv T_{\mu +1}(k)/T_1(k)\qquad \qquad (\mu =1,2,3)  \label{f81}
\end{equation}

\noindent Clearly, $\left\{ F^2\right\} $ is identical to $\bf{f}^T%
\bf{f}.$

This simple and elegant expression for $R(k)$ arises in a quite general way
from the dyadic structure of $\widetilde{q}_{ij}(k)$ (only some properties
of $\alpha _j$ were added, i.e., $\alpha _j(-k)=\alpha _j(k)=\alpha
_j^{*}(k)\ $). Of course, such a formula might be used as is. Nevertheless,
from a practical point of view, it is more convenient to take advantage of
the particular, explicit, expressions for $\alpha _{j,}\ \beta _j$ and $%
\gamma _j$ in the present model, in order to achieve a sort of ``reduction''
of the above-written result. Since simpler expressions are preferable for
numerical calculations, we have tried to perform all possible
simplifications, starting from the determinant form of $L_j\left( k\right) $
and using some properties of determinants.

The first step is to take the determinant form of $L_j\left( k\right) $, eq.
(\ref{f77}), and subtract from the third column the second one multiplied by 
$\left( 3\xi _2/\Delta -ik/2\right) .$ Since $\beta _{1,\ j}$ is given by
eq. (\ref{f59}), such on operation ``eliminates'' $\beta _{1,\ j}$ from the
third column, which becomes

\begin{equation}
\begin{array}{l}
\beta _{0,\ j} \\ 
\\ 
\left\langle \beta _0\right\rangle -\frac{3\xi _2}\Delta +\frac 12ik \\ 
\\ 
~1+\left\langle \sigma \beta _0\right\rangle  \\ 
\\ 
\left\langle A\beta _0\right\rangle 
\end{array}
\label{f82}
\end{equation}

\noindent The advantage of the resulting expression for $L_j\left( k\right) $
is that, in place of the complex functions $\beta _j$, it involves only the $%
\beta _{0,\ j}$, which are real and even. It is worth noting that such a
reduction is implicit in Vrij's work \cite{Vrij79}.

Let us now transform the electrostatic contributions. Our second step
``eliminates'' $\gamma _{2,\ j}$ from the fourth column. Looking at eq. (\ref
{f63}), it is evident that it is now necessary to subract a linear
combination of the second column and the third one (in the new form with $%
\beta _{0,\ j}$), multiplied respectively by $P_z\ \Delta $ and $2iP_z\
\Delta \ /k.$ The fourth column becomes

\begin{equation}
\begin{array}{l}
\gamma _{01,\ j}\equiv \gamma _{0,\ j}+\gamma _{1,\ j} \\ 
\\ 
\left\langle \gamma _{01}\right\rangle +\frac{2i}k3\xi _2P_z \\
\\ 
~\left\langle \sigma \gamma _{01}\right\rangle -\frac{2i}kP_z\Delta  \\
\\ 
1+\left\langle A\gamma _{01}\right\rangle 
\end{array}
\label{f83}
\end{equation}

As a further step, we have observed that the particular functional form of $%
\gamma _{1,\ j}(k)$ allows a direct analytic evaluation of the necessary
averages relevant to this term. By deriving from eq.s (\ref{f44}),(\ref{f48}%
), (\ref{f49}) and from the electroneutrality condition the following new
identities 

\begin{equation}
\left\{ A\right\} \equiv \sum_m\rho _mA_m=-\frac{\ell _B}{\pi \Gamma }%
P_z\left( 3\xi _2+\Gamma \Delta \right)   \label{f84}
\end{equation}

\begin{equation}
\left\{ \sigma A\right\} \equiv \sum_m\rho _m\sigma _mA_m=\frac{\ell _B}%
\Gamma P_z\Delta ,  \label{f85}
\end{equation}

\noindent together with eq. (\ref{f51}) for $\left\{ A^2\right\} \equiv
\sum_m\rho _mA_m^2=\ell _B/\pi \ $, we find

\begin{equation}
\left\langle \gamma _1\right\rangle =-i\frac 2kP_z\left( 3\xi _2+\Gamma
\Delta \right)   \label{f86}
\end{equation}

\begin{equation}
\left\langle \sigma \gamma _1\right\rangle =i\frac 2kP_z\Delta   \label{f87}
\end{equation}

\begin{equation}
\left\langle A\gamma _1\right\rangle =i\frac{2\Gamma }k  \label{f88}
\end{equation}

Putting all these results together and multiplying $L_j(k)$ by $k$ to
``regularize'' its behaviour at $k=0$ (as proposed in eq. (\ref{f74}) we get

\begin{equation}
kL_j(k)=\rho _j^{1/2}\left| 
\begin{array}{llll}
~F_j & \qquad \alpha _j & \ \qquad \beta _{0,\ j} & \quad \quad k\gamma
_{01,\ j} \\ 
\left\langle F\right\rangle  & \quad 1+\ \left\langle \alpha \right\rangle 
& ~\left\langle \beta _0\right\rangle -3\xi _2/\Delta +ik/2 & \quad
k\left\langle \gamma _0\right\rangle -2\pi iP_z\Gamma \Delta  \\ 
\left\langle \sigma F\right\rangle  & \qquad \left\langle \sigma \alpha
\right\rangle  & \quad 1+\left\langle \sigma \beta _0\right\rangle  & \qquad
k\left\langle \sigma \gamma _0\right\rangle  \\ 
\left\langle AF\right\rangle  & \qquad \left\langle A\alpha \right\rangle  & 
\qquad \left\langle A\beta _0\right\rangle  & \quad k\left[ 1+\left\langle
A\gamma _0\right\rangle \right] +i2\Gamma
\end{array}
\right|   \label{f89}
\end{equation}

\noindent where

\begin{equation}
\gamma _{01,\ j}(k)\equiv \gamma _{0,\ j}(k)+\gamma _{1,\ j}(k)=\frac{\Gamma
A_j}{\ell _B}\ \frac{2\pi }k\left[ \sin X_j+i\left( \cos X_j+\frac{2\Gamma }%
k\sin X_j\right) \right]   \label{f90}
\end{equation}

This new form for $kL_j(k)$ is really more convenient than that
corresponding to eq. (\ref{f77}), since it requires a smaller number of
numerical calculations and allows to write a simpler expression for the
scattering intensity. In fact, expanding the aforesaid determinant along the
first line and taking into account that $F_j,\ \alpha _j$ and $\beta _{0,\ j}
$ are real and even functions, we get after some algebra the following final
result for $R(k)$

\begin{equation}
R(k)=R_1(k)+R_2(k),  \label{f91}
\end{equation}

\begin{eqnarray}
R_1(k) &=&\left\{ F^2\right\} +\left\{ \alpha ^2\right\} C_1C_1^{*}+\left\{
\beta _0^2\right\} C_2C_2^{*}  \label{f92} \\
&&+2\mbox{Re}\left[ \left\{ F\alpha \right\} C_1+\left\{ F\beta _0\right\}
C_2+\left\{ \alpha \beta _0\right\} C_1C_2^{*}\right]   \nonumber
\end{eqnarray}

\begin{eqnarray}
R_2(k) &=&k^2\left\{ \gamma _{01}\gamma _{01}^{*}\right\} C_3C_3^{*}
\label{f93} \\
&&+2k\ \mbox{Re}\left[ \left\{ \gamma _{01}F\right\} C_3+\left\{ \gamma
_{01}\alpha \right\} C_3C_1^{*}+\left\{ \gamma _{01}\beta _0\right\}
C_3C_2^{*}\right]   \nonumber
\end{eqnarray}

 \noindent where $\mbox{Re}\left[ ...\right] $ denotes the real part of a
 complex value function and the $C_{\mu} $ used here are not the same as
 those in eq. (\ref{f79}): their definition is again given by eq. (\ref{f81})
 but the cofactors $T_{\mu}(k)$ now refer to the new determinant expression
 for $kL_j(k)$, i.e., to eq. (\ref{f89}). Note that $T_1(k)=kD_Q(k)$.

Observe that $\left\{ F\alpha \right\} ,\ \left\{ F\beta _0\right\} $ and $%
\left\{ \alpha \beta _0\right\} $ are real quantities, while the $\gamma
_{0,\ j}(k)$ are odd and purely immaginary functions, i.e., $\gamma _{0,\
j}(-k)=-\gamma _{0,\ j}(k)=\gamma _{0,\ j}^{*}(k).$

In the limit of $\left\{ \mbox{all charges }z_m\rightarrow 0\right\} $, it
results that $R_2(k)\rightarrow 0$ and $R_1(k)\rightarrow R^{HS}(k)$, i.e.,
we recover exactly (apart from the notation) Vrij's formula for the neutral
HS case.

Finally, as concerns the partial structure factors, they can easily be
evaluated by means of eq.s (\ref{f16}) and (\ref{f69})-(\ref{f71}), but we
prefer not to give their explicit expressions here.

\section{The Polydisperse Limit}

In the treatment of polydisperse systems it is customary to start from a
mixture with a finite number of components and then to perform a limiting
process, which results into a continuous distribution of the properties
which characterize the species ({\it identity variables}) \cite
{Gualtieri82}. In other words, in an IE approach one first solves the
equations and then takes the infinite-species limit. This is a delicate
point.

Still in the IE framework, we like to emphasize a second alternative route,
i.e., the possibility of taking the infinite-species limit from the outset.
According to this second viewpoint; one first sets up integral equations for
polydisperse fluids and then tries to solve them. It is worth mentioning
that, although a polydisperse version of the OZ equations for HS has already
appeared in the literature \cite{Mcrae88}, we are unaware of any papers
where polydisperse Baxter integral equations have been proposed.

To this aim, let us first recall how the composition of polydisperse systems
can be specified. In a ``fully polydisperse'' mixture the discrete variable $%
1\leq i\leq p$ that labels the species is replaced by a set of one, two or
more continuous variables, which constitute an {\it identity vector }$%
\bf{I}=\left\{ I_1,I_2,...,I_s\right\} $ (not to be confused with the
previously used identity matrix). The identity variables are distributed
according to a multivariable {\it molar fraction density function }$p(%
\bf{I)}$ \cite{Gualtieri82}, \cite{Briano83}, \cite{Kofke86}: the molar
fraction $x_i=\rho _i/\rho =\rho _i/\sum_m\rho _m$ of the discrete case goes
into

\begin{equation}
p(\bf{I)}d {\bf I}=p(I_1,I_2,...,I_s) dI_1dI_2...dI_s,
\label{f94}
\end{equation}

\noindent which is the probability of finding a particle with species labels
in the identity range $(\bf{I,I+}d {\bf I).}$ As a consequence, for
the number density it results that

\begin{equation}
\rho _i\rightarrow \rho p({\bf I)}d {\bf I},  \label{f95}
\end{equation}

\noindent with $\rho $ being the total number density.

In the opposite case of  ``fully monodisperse'' mixtures, one could write

\begin{equation}
p(\bf{I)=}\sum_{\left\{ {\bf I}^{\prime }\right\} }x\left( {\bf I}%
^{\prime }\right) \delta \left( \bf{I-I}^{\prime }\right) ,  \label{f96}
\end{equation}

\noindent where $\left\{ \bf{I}^{\prime }\right\} $ is the set of
identity vectors corresponding to discrete species present in the finite
amounts $\left\{ x\left( \bf{I}^{\prime }\right) \right\} $ (here, $%
\delta $ is the Dirac delta distribution).

Of course, one may also consider ``partly polydisperse'' mixtures in which
several species are monodisperse, while the remaining ones are polydisperse.
In this case, the distribution function $p(\bf{I)}$ is the sum of a
discrete part, $p_d(\bf{I)}$, like that in eq. (\ref{f96}) and a
continuous one, $p_c(\bf{I)}$. Clearly, to be a probability density
function, $p(\bf{I)}$ must satisfy the normalization condition

\begin{equation}
\sum_{\left\{ \bf{I}^{\prime }\right\} }x\left( {\bf I}^{\prime
}\right) +\int d\bf{I\ }p_c({\bf I)}=1  \label{f97}
\end{equation}

Bearing all this in mind, it is immediate to pass from the Baxter equations (%
\ref{f37}) for $p$-component mixtures to those for polydisperse fluids,
using the following the replacement

\begin{equation}
\sum_m\rho _m\ ...\rightarrow \int d\rho _m\ ...=\rho \int d\bf{I\ }p(%
\bf{I)\ ...,}  \label{f98}
\end{equation}

\noindent if we re-interpret the species label $m$ as a shorthand notation
for the identity vector $\bf{I,}$ whose distribution function $p({\bf
I)}$ may include both a continuous and a discrete part, respectively for
polydisperse and monodisperse components.

Using this simple convention, the procedure of solving the polydisperse
Baxter equations for uncharged or charges HS, within the MSA, is formally
the same as in the monodisperse case, apart from the substitution required
by eq. (\ref{f98}). In other words, we can conclude that, at least for the
two aforesaid systems, the MSA solution for the polydisperse case can be
obtained from the monodisperse one via the simple prescription expressed by
eq. (\ref{f98}). Note that all this fully agrees with the findings by Stell
and coworkers \cite{Stell79}, \cite{Salacuse82} for neutral HS in the PY
approximation and with the MSA investigation performed by Senatore and Blum 
\cite{Senatore85} on mixtures of charged HS with either size polydispersity
(and fixed charge value) or charge polydispersity (and fixed diameter).

Before concluding this Section, it is to be noted that, while for HS a
single identity variable\ - \ the diameter\ -\ is sufficient to characterize
a species (i.e., $\bf{I}=\left\{ \sigma \right\} $), for charged HS two
variables\ - \ the diameter and the charge\ -\ are necessary (i.e., $\bf{%
I}=\left\{ \sigma ,z\right\} $). To avoid the difficulty of working with a
two-variable distribution $p\left( \sigma ,z\right) ,$ one might assume a
strong correlation between $\sigma $ and $z,$ reducing the number of
independent variables to only one. It is customary to express such a
relationship in the form $z=z\left( \sigma \right) $ \cite{Senatore85}, 
\cite{Lowen91},
 but perhaps the inverse function $\sigma =\sigma \left( z\right) $
might be preferred in view of the integrations to be done over the remaining
independent variable (recall that the range of $\sigma $ is $\left[
0,+\infty \right[ $, while $z$ varies in the interval $\left] -\infty
,+\infty \right[ \ $). Anyway, for macroions of colloidal suspensions it is
reasonable to assume that their charges scale linearly with their surface
area, i.e., $z\propto \sigma ^2$, keeping the surface charge density
constant \cite{Aguan90}, \cite{Aguan91}, \cite{Aguan92}.

\section{Concluding Remarks}

The first and main result of this paper is given by eq.s (\ref{f91})-(\ref
{f93}), which provide a new MSA formula for the scattering intensity from
multicomponent mixtures of charged hard spheres. This fills an important gap
present so far in the relevant literature. It can be applied to a wide
variety of ionic fluids and is expected to be particularly useful to fit
experimental data.

Our result extends Vrij's one for neutral HS and reduces to it in the limit
of vanishing electrostatic contributions. Despite the presence of the
electrostatic terms, this case is still analytically tractable, due to the
dyadic form of $\widetilde{q}_{ij}(k).$ Indeed, the rank of the matrices
involved in such calculations turns out to be independent of the number $p$
of components in the mixture, always being equal to the number $n$
(typically $\ll p$) of dyads constituting the $n$-dyadics $\widetilde{q}%
_{ij}(k).$ Such a feature allows to take the polydisperse $p\rightarrow
\infty $ limit with an arbitrary continuous distribution of both charges and
sizes.

Our findings are by no means academics. The importance of having an
analytical expression of the scattering intensity for experimental purposes
has been well established \cite{Abis90}. An application of the present
results to neutron scattering experimental data will be the subject of a
forthcoming paper.

A word of caution on the range of applicability of our results is in order.

The failure of the MSA closure at low densities and high charge values could
be overcome by using a rescaling procedure (RMSA) \cite{Hansen82}, \cite
{Belloni86}, \cite{Ruiz90}. However, this does not present a problem for
many ionic fluids of current interest, which are made up of weakly charged
particles at high volume fractions. In such cases the MSA is expected to be
a reasonably accurate approximation, yielding semiquantitatively good
predictions for the static structure factors.

Regarding the non-linear equation for the single parameter $\Gamma $, which
has to be solved in the MSA scheme, selecting the physical root from the
manifold of solutions is not a terribly difficult task \cite{Pastore88}. In
fact, as Hiroike \cite{Hiroike77} pointed out, this equation has only one
real positive solution which is the physical one. Moreover, in their
application of the MSA for charged HS to micellar solutions Senatore and
Blum \cite{Senatore85} already solved the equation for $\Gamma $ even in the
polydisperse limit without mentioning any particular problems.

As far as the potential considered in this work and the possibility of
extending our MSA scheme to other different interactions are concerned, it
is evident that the primitive model represents the simplest physically
significant choice to describe ionic fluid mixtures.

The primitive model includes both attractive and repulsive Coulomb
potentials and has the advantage of treating large and small ions on the
same footing. For colloidal suspensions, it definitely gives a better
description than the essentially one-component models which consider only
the repulsive (screened) Coulomb interactions between macroions, while the
other species (counter-ions, any added electrolyte and solvent) are taken
into account only upon determining the screening in the effective
macroion-macroion potential \cite{Hayter81}, \cite{Hansen82}, \cite{Wagner91}%
.

Of course, several alternative choices of potential model may be found in
the literature on scattering from multicomponent fluids. For instance, very
recently, Kline and Kaler \cite{Kline96} fitted their experimental partial
structure factors of a mixed colloidal system (sodium dodecyl sulfate
micelles plus Ludox colloidal silica) by a model of hard and sticky hard
spheres (in the terminology of these authors, ``sticky'' means that an
attractive square well is added outside the hard core).

Indubitably the next step after the primitive model should be the mixture of
particles interacting by HS plus Yukawa ($\equiv \ $screened Coulomb)
potentials, i.e., $\beta \phi _{ij}(r)=K_{ij}\exp (-\mu r)/r$ for $r>\sigma
_{ij}$. Unfortunately, the general MSA solution for multicomponent Yukawa
fluids \cite{Blum78}, \cite{Blum80B} is much more involved than the MSA one
for unscreened charged HS, even if substantial simplifications occur for
factorizable coupling parameters, i.e., when $K_{ij}=K_iK_j$ \cite{Ginoza86}%
, \cite{Herrera96}. In particular, a satisfactory treatment of polydisperse
Yukawa systems is still lacking. In this context, we recall that L\"{o}wen
et al. \cite{Lowen91} proposed a mapping of the polydisperse Yukawa
model onto the polydisperse HS reference system. It is also worth mentioning
that, to avoid the difficulty of working with the MSA solution for Yukawa
mixtures, Ruiz-Estrada et al. \cite{Ruiz90} invented a method to
map a Yukawa system onto an equivalent (but much easier to be treated)
primitive model system. Very recently, an analytical equation of state for
the HS Yukawa polydisperse fluid (in the equal-diameter case with
polydispersity in the coupling parameters $K_i$) has been presented \cite
{Ginoza97}.We believe that the Yukawa model deserves further investigation,
specially on the possibility of deriving a closed MSA formula for the
scattering intensity in this case as well. We are planning to do it in the
near future.

Finally, before concluding this paper, we remark that our methodological
recipe, based on the properties of the dyadic matrices presented in Section
III, has far reaching consequences. It is indeed the property of
``contraction'' or ``reduction'' of these matrices (according to which $%
\left| \widehat{\bf{Q}}(k)\right| $ can be reduced to an equivalent
determinant of very low order) that makes the evaluation of the structure
factors for fluids with a large or infinite number of components a tractable
problem. Our findings have also led to a simple recipe for building the
inverse matrix $\widehat{\bf{Q}}^{-1}(k)$ for the most general $%
\widetilde{q}_{ij}(k)$ of dyadic form. Our formula not only includes those
previously found by Blum and coworkers for $n=2$ and $n=3$ \cite{Stell79}, 
\cite{Blum77}, but, more importantly, it provides a systematic way to extend
the results of the present paper to new more complex cases with $n>3.$

\vskip 2cm

\acknowledgments
Partial financial support by the Italian MURST (Ministero
dell'Universit\`{a} e della Ricerca Scientifica through the INFM (Istituto
Nazionale di Fisica della Materia) is gratefully acknowledged. One of us
(AG) is grateful to Prof. Klaus Kehr for his hospitality at the FZJ during
the initial part of this work.

\newpage

\appendix

\section{Proofs for Section III}

I) The proof of (\ref{f26}) is patterned after a similar one used in the
Laplace expansion of the characteristic polynomial of a $p\times p$ matrix.

Consider a $p\times p$ matrix of the generic form: 
\begin{equation}
M_{ij}=\delta _{ij}+\widehat{W}_{ij}=\delta _{ij}+(\rho _i\rho
_j)^{1/2}W_{ij}    \label{a1}
\end{equation}

For the sake of simplicity, we shall work out the case $p=3$ first and then
show how the result can be generalized. Let us rewrite the determinant of
the $3\times 3$ matrix (A1) in such a way that each element is the sum
of two terms (type A and type B respectively)

\begin{equation}
|\bf{M}|=\left| 
\begin{array}{ccc}
1+\rho _1W_{11} & \quad 0+(\rho _1\rho _2)^{1/2}W_{12} & \quad 0+(\rho
_1\rho _3)^{1/2}W_{13} \\ 
0+(\rho _2\rho _1)^{1/2}W_{21} & 1+\rho _2W_{22} & \quad 0+(\rho _2\rho
_3)^{1/2}W_{23} \\ 
0+(\rho _3\rho _1)^{1/2}W_{31} & \quad 0+(\rho _3\rho _2)^{1/2}W_{32} & 
1+\rho _3W_{33}
\end{array}
\right|    \label{a2}
\end{equation}

Using elementary properties of determinants, the above expression can be
written as a sum of a determinant containing only elements of type A (i.e.,
the identity matrix), three determinants containing two rows of type A and
one of type B, three determinants containing one row of type A and two of
type B and, finally, one determinant containing only rows of the type B.
Hence there are $2^3$ determinants overall. Again exploiting elementary
properties of determinants and symmetry considerations on the permutation of
the indices, one easily establishes that (A2) can be written as

\begin{equation}
|\bf{M}|=1+\sum_{i=1}^3\rho _iW_{ii}+\frac 1{2!}\sum_{i,j=1}^3\left| 
\begin{array}{ll}
W_{ii} & W_{ij} \\ 
W_{ji} & W_{jj}
\end{array}
\right| +\rho _1\rho _2\rho _3\left| 
\begin{array}{ccc}
W_{11} & W_{12} & W_{13} \\ 
W_{21} & W_{22} & W_{23} \\ 
W_{31} & W_{32} & W_{33}
\end{array}
\right|   
\end{equation}

The above expression can be generalized to the general $p$-component case,
by noting that all terms with $s<p$ rows of type A and $p-s$ of type B can
be written as

\begin{equation}
\frac 1{(p-s-1)!}\sum_{j_1,..,j_{p-s-1}=1}^p\rho _1\dots \rho _{p-s-1}\left| 
\begin{array}{ccc}
W_{s+1,s+1} & \dots  & W_{s+1,p} \\ 
&  &  \\ 
\vdots  & \ddots  & \vdots  \\ 
W_{p,s+1} & \dots  & W_{p,p}
\end{array}
\right|   
\end{equation}

Collecting terms of all orders yields then (\ref{f26}).

II) Let us now show that, if $\widehat{W}_{ij}$ is an $n$-dyadic (with $n<p$%
), then any minor of $\widehat{\bf{W}}$ having order $m>n$ vanishes.

We outline the proof only for $n=2,$ since its generalization to $n>2$ is
very easy. We have to show that any minor of order $m\geq 3$ is zero, but,
since all minors of order $m>3$ can be expressed in terms of the third order
ones, it is sufficient to demonstrate that

\begin{equation}
\left| 
\begin{array}{ccc}
\widehat{W}_{iq} & \widehat{W}_{ir} & \widehat{W}_{is} \\ 
\widehat{W}_{jq} & \widehat{W}_{jr} & \widehat{W}_{js} \\ 
\widehat{W}_{kq} & \widehat{W}_{kr} & \widehat{W}_{ks}
\end{array}
\right| =0\qquad \forall (i,j,k,q,r,s)    \label{a5}
\end{equation}

\noindent If $\widehat{W}_{ij}=a_i^{(1)}b_j^{(1)}+a_i^{(2)}b_j^{(2)}$ and if
we define

\begin{equation}
\lambda _i\equiv a_i^{(2)}/a_i^{(1)}\qquad \mbox{and \qquad }\mu _j\equiv
b_j^{(2)}/b_j^{(1)}\qquad \mbox{(if \quad }a_i^{(1)}\neq 0,\ b_j^{(1)}\neq 0%
\mbox{),}  
\end{equation}

\noindent then we could write $\widehat{W}_{ij}=a_i^{(1)}b_j^{(1)}\left(
1+\lambda _i\mu _j\right) $. Substituting this expression into the
determinant of eq. (A5) and factoring out all common factors, we get

\begin{equation}
a_i^{(1)}a_j^{(1)}a_k^{(1)}b_i^{(1)}b_j^{(1)}b_k^{(1)}\left| 
\begin{array}{ccc}
1+\lambda _i\mu _q & \quad 1+\lambda _i\mu _r & \quad 1+\lambda _i\mu _s \\ 
1+\lambda _j\mu _q & \quad 1+\lambda _j\mu _r & \quad 1+\lambda _j\mu _s \\ 
1+\lambda _k\mu _q & \quad 1+\lambda _k\mu _r & \quad 1+\lambda _k\mu _s
\end{array}
\right|   
\end{equation}

\noindent Then, by subtracting the third row from both the first and the
second one, we see that the resulting determinant vanishes since two rows
are proportional.

III) As concerns $|\bf{M}|$ expressed in terms of $\delta _{\mu \nu }+%
\bf{a}^{(\mu )}\cdot {\bf b}^{(\nu )},$ eq. (\ref{f28}) can be
verified very rapidly for $n=2$, by using eq.s (\ref{f26}) and (A5) together
with the following identity

\begin{equation}
\left| 
\begin{array}{cc}
\widehat{W}_{iq} & \widehat{W}_{ir} \\ 
\widehat{W}_{jq} & \widehat{W}_{jr}
\end{array}
\right| =\left[ a_i^{(1)}a_j^{(2)}-a_i^{(2)}a_j^{(1)}\right] \left[
b_q^{(1)}b_r^{(2)}-b_q^{(2)}b_r^{(1)}\right] , 
\end{equation}

\noindent which holds true for 2-dyadics.

IV) Finally, in order to prove eq.s (\ref{f30})-(\ref{f32}) for the
cofactors, one first establishes that the elements $M_{ij}^{-1}$ of the
inverse matrix can be computed order by order in the $\rho $'s, by using the
definition

\begin{equation}
\sum_{k=1}^pM_{ik}^{-1}M_{kj}=\delta _{ij}  
\end{equation}

\noindent This leads to the following result

\begin{equation}
M_{ij}^{-1}=\delta _{ij}-(\rho _i\rho _j)^{1/2}W_{ij}+(\rho _i\rho
_j)^{1/2}\sum_{k=1}^p\rho _kW_{ik}W_{kj}-(\rho _i\rho
_j)^{1/2}\sum_{k,l=1}^p\rho _k\rho _lW_{ik}W_{kl}W_{lj}+...  
\end{equation}

\noindent Then, starting from $|\bf{M}|^{ji}=M_{ij}^{-1}| {\bf M}|$,
using the expansion (\ref{f26}) for $|\bf{M}|$ and collecting all terms
of the same order, one finds, after some simple algebra, the required
expansion.

\section{Comparison between different notations}

In place of Baxter's factorization expressed by eq. (\ref{f13}), Blum and
coworkers \cite{Stell79}, \cite{Blum75}, \cite{Blum77}, \cite{Blum78} start
from

\begin{equation}
\bf{I}-{\bf C}(k)=\widehat{{\bf Q}}\left( k\right) \widehat{%
\bf{Q}}^T\left( -k\right)     \label{b1}
\end{equation}

\noindent where the elements of $~\widehat{\bf{Q}}\left( k\right) $ are
again of the form 
\begin{equation}
\widehat{\bf{Q}}\left( k\right) ={\bf I}-\widetilde{{\bf Q}}\left(
k\right) ,    \label{b2}
\end{equation}

\noindent but with

\begin{equation}
\widetilde{\bf{Q}}\left( k\right) =2\pi \int_{-\infty }^{+\infty }dr\
e^{ikr}\bf{Q}\left( r\right) \qquad \mbox{and \qquad }Q_{ij}(r)=\left(
\rho _i\rho _j\right) ^{1/2}q_{ij}(r)    \label{b3}
\end{equation}

This choice leads to a slightly different version of the equations for the
factor correlation functions, namely to

\begin{equation}
\left\{ 
\begin{array}{l}
2\pi rc_{ij}\left( |r|\right) =-q_{ij}^{\ \prime }(r)+\sum_m\rho
_m\int_{\lambda_{mj}}^\infty dt\ q_{jm}\left( t\right) q_{im}^{\ \prime }\left(
r+t\right) ,\qquad r>\lambda _{ji}\equiv \frac 12\left( \sigma _j-\sigma
_i\right)  \\ 
\\ 
2\pi rh_{ij}\left( |r|\right) =-q_{ij}^{\ \prime }(r)+2\pi \sum_m\rho
_m\int_{\lambda _{jm}}^\infty dt\ \left( r-t\right) h_{im}\left(
|r-t|\right) q_{mj}\left( t\right) ,\qquad r>\lambda _{ji}
\end{array}
\right.     \label{b4}
\end{equation}

The correspondence between our (or Baxter's) notation and Blum's one is very
simple

\begin{equation}
q_{ij}^{Baxter}(r)=\frac 1{2\pi }q_{ji}^{Blum}(r)  
\end{equation}

\noindent As a consequence, our $q_{ij}^{\ \prime },$ $q_i^{\ \prime \prime }
$ and $A_i$ are easily obtained from Blum's counterparts by performing
index transposition and division by $2\pi .$

Finally, our $P_z$ and $\Omega $ correspond, respectively, to $\Delta
^{-1}P_N$ (or $\Delta ^{-1}P_n)$ and $\Omega \Delta $ of Blum's papers.

\vskip 2cm


\end{document}